\documentclass[twocolumn,nofootinbib]{revtex4-1}
\pdfoutput=1

\usepackage{graphicx,amsmath,amssymb}

\usepackage{bbm,bm,amsmath,amssymb}

\def\bg{\begin{eqnarray}}
\def\nd{\end{eqnarray}}

\def\cos{{\rm cos}}

\def\log{{\rm log}}

\begin{document}

\title{Axion Hilltops, Kahler Modulus Quintessence and the Swampland Criteria}
\author{Maxim Emelin}
\affiliation{Department of Physics, McGill University, Montr\'{e}al,  Qu\'{e}bec,  H3A 2T8, Canada}
\email{maxim.emelin@mail.mcgill.ca}
\author{Radu Tatar}
\affiliation{Department of Mathematical Sciences, University of Liverpool, Liverpool,  L69 7ZL, United Kingdom}
\email{Radu.Tatar@Liverpool.ac.uk}

\begin{abstract}We study the interplay between extrema of axion potentials, Kahler moduli stabilization and the swampland criteria. We argue that moving away from the minima of non-perturbatively generated axion potentials can lead to a runaway behavior of moduli that govern the couplings in the effective field theory. The proper inclusion of these degrees of freedom resolves the conflict between periodic axion potentials and the gradient de Sitter criterion, without the need to invoke the refined de Sitter criterion. We investigate the possibility of including this runaway direction as a model of quintessence that satisfies the swampland criteria. Using a single non-perturbative effect, the axionic maximum provides a runaway direction that is unstable in the axion directions, sensitive to initial conditions and too steep to allow for a Hubble time of expansion without violating the field excursion criterion. Adding a second non-perturbative effect generates a saddle point in the potential, which solves the steepness problem and improves the initial conditions problem although some fine-tuning remains required.

\end{abstract}

\maketitle

\section{Introduction}

Axions are ubiquitous in theoretical physics. The Peccei-Quinn mechanism \cite{pq}, a proposed resolution to the strong CP problem, promotes the QCD $\theta$-angle to a dynamical axion field. Multi-axion models are also commonly used for cosmological model building, for example natural inflation \cite{natinflation} models, axion quintessence models \cite{aquint} etc. Axions are also omnipresent in string theory compactifications. In type II strings they arise from the Ramond-Ramond p-form potentials wrapping internal cycles of the compactification manifold and in the case of SUSY compactifications, give the imaginary parts of the chiral multiplets in the 4D effective theory \cite{witten}. 

In this paper we explore various aspects of axion models that originate in string theory and their interplay with the swampland criteria \cite{swampland} that received much attention over the past few months (e.g. \cite{murayama, higgs, phenomquint, others, moritz, obied, wachter, susha} among many others). Our particular focus will be on the interplay of axions in string theory with what we'll refer to as the ``distance'', ``gradient'' and ``hessian'' criteria. The distance criterion states that field values must remain within a planck unit of the ground state of our effective field theory. The gradient criterion states that the scalar potential must obey 
\bg \label{swampland}
\nabla V \geq c V  ~~~~~~~~~~ \tilde{c} = \mathcal{O}(1).
\nd 

This criterion has been refined, as proposed in \cite{krishnan} and further justified in \cite{swampland2} (see also \cite{swamplandrefine}), to include a condition on the second derivative. This refined criterion allows for violations of the gradient criterion, provided the Hessian of the potential has a sufficiently negative eigenvalue. 

\bg
\min \nabla_\mu \nabla_\nu V \leq a V ~~~~~~~~ \tilde{a} = \mathcal{O}(1)
\nd

This is the hessian criterion. In this paper we will try to avoid making use of the hessian criterion, although we will be naturally led into making use of it in the second half of section \ref{ht} 

In section \ref{pq} we point out that the hilltops of axion potentials are not in conflict with the gradient de Sitter swampland criterion \cite{swampland} in the most natural embeddings of axions in string theory. The reason is that the same non-perturbative effects that generate the axion potentials also generate a potential for the moduli that govern the coupling and the total potential satisfies the gradient criterion. We do not touch the similar apparent conflict in the case of the Higgs potential \cite{murayama, higgs} as the embeddings of the Higgs potential into a string theoretical setup are less clear.

In section \ref{ht} we attempt to harness this non-perturbatively generated potential to construct a quintessence model. This model uses the real part of a Kahler modulus at an axionic hilltop as a quintessence field. We forego the use of any supersymmetry breaking uplift ingredient such as anti-D3 branes. We will find that the simplest such model with one non-perturbative contribution to the superpotential suffers from several issues. The model requires extreme fine-tuning of a one-loop determinant coefficient to ensure the smallness of the potential as well as fine tuning of initial conditions. Furthermore, we find that for all values of the parameters, the slow-roll parameter is bounded below and is in tension with observations as well as the distance criterion. We attempt to remedy the initial conditions problem by considering additional non-perturbative effects in a racetrack scenario, and find that this also naturally solves the slow-roll problem and produces a thawing quintessence model. We conclude with a review of our results as well as some general comments and discussion. \footnote{
As this work was being completed papers involving a similar set of ingredients appeared \cite{obied, wachter, susha}. The overlap with \cite{wachter} concerns discussions of transplanckian axions, \cite{susha} also considers using the size modulus as a thawing hilltop quintessence, but without axion dynamics and no tree level superpotential. \cite{obied} contains a useful analysis of the constraints on initial conditions near hilltops, which we incorporated in section \ref{ht}.}

\section{Peccei-Quinn in String Theory} \label{pq}

It was recently suggested \cite{murayama} that the Peccei-Quinn mechanism for solving the strong CP problem is in tension with the gradient criterion \eqref{swampland}.

The authors of \cite{murayama} consider a simple model of an action with a potential consisting of a small slowly varying quintessence term and the usual cosine potential for the QCD axion field, with parameters within the bounds dictated by observation. They then observed that at the local maximum of the cosine potential the swampland criterion \eqref{swampland} is violated by several orders of magnitude. A modified criterion has been proposed in \cite{swampland2}, which among other things removes this conflict, however we will argue that the old criterion is also not violated if the theory is properly embedded in string theory. 

Here we analyze realizations of this scenario in string theory and show how this problem is averted. The key point is that the gauge coupling is itself a dynamical degree of freedom in string theory and the same non-perturbative effects that generate the axion potential also couple the potential to the gauge coupling.

The Peccei-Quinn mechanism \cite{pq} for the resolution of the strong CP problem instructs us to promote the QCD $\theta$-angle to a dynamical degree of freedom $a$. This degree of freedom naturally couples to the Yang-Mills instanton density $F\wedge F$ and the effect of the instantons generates an effective potential for $a$. This potential can be approximated by a dilute instanton gas calculation \cite{coleman}. The theory has approximate saddles corresponding to $n$ instantons and $\bar{n}$ anti-instantons, all widely separated. Evaluating the contribution of these saddles to the path integral gives

\bg
\sum_{n,\bar{n}}\frac{1}{n! \bar{n}!}& (K e^{-S_0})^{n+\bar{n}} e^{i(n-\bar{n}) a} \mathcal{V}^{n+\bar{n}} = \nonumber \\
&exp(2 K \mathcal{V} e^{-S_0} \cos a),
\nd
where $S_0$ is the single-instanton action, $K$ is a one-loop determinant around a single instanton and $\mathcal{V}$ is the volume of the moduli space of a single instanton. This volume is proportional to the volume of the spacetime, since there is a zero-mode corresponding to the location of the instanton, however in the presence of internal symmetries (e.g. R-symmetry) there are additional moduli that need to be integrated over, giving additional multiplicative factors.

More importantly, the action of the instanton itself depends on the value of other moduli, specifically the dilaton and the volumes of internal cycles.

Let's consider some specific examples. We start with a stack of D3 branes in a type IIB compactification. The effective 4D theory will of course contain the $\mathcal{N}=4$ SYM action coming from the worldvolume of the branes, but since the internal space is compact it will also include terms governing the dynamics of all the moduli fields that aren't stabilized at energies above the cutoff. The worldvolume action for the branes contains a coupling to the RR axion:

\bg
\int d^4 x C_0 F\wedge F
\nd
so $C_0$ is precisely the $\theta$-angle in the $\mathcal{N}=4$ SYM, but in the full theory it is in fact a dynamical field. We can now compute the non-perturbative potential induced by the instantons.

The instanton action is proportional to the inverse string coupling
\bg
S_0 \propto \frac{1}{g_{YM}^2}= \frac{1}{g_s}=e^{-\phi}
\nd

In fact we can recognize the Yang-Mills instantons as D-instantons dissolved inside the branes. In the absence of warping, there are also contributions from D-instantons located at any point in the internal space that couple to $C_0$ in exactly the same way. The full non-perturbative potential for $C_0$ will then include an integral over the whole moduli space of these D-instantons giving an overall factor of the internal volume. The leading order non-perturbative contribution to the potential coming from the instantons is 
\bg \label{ep}
 V \propto - A e^{-S_0} \cos \,a = - A \exp(\alpha e^{-\phi})  \, \cos \, a,
\nd

with $A$ proportional to the volume of the instanton moduli space as well as the one-loop determinant around a single instanton.

We now see that the potential conflict with the swampland criterion \eqref{swampland} due to the local maximum at $a = \pi$ is alleviated for small values of $\phi$, i.e. at weak coupling. If we fix the axion to be at the local maximum of the cosine, the potential still has a dependence on the dilaton, which satisfies the swampland criterion at weak coupling.

We can consider alternative realizations of axions. For example we can consider higher dimensional branes wrapping internal cycles. Consider a D5 wrapping a 2-cycle. Then similarly to the D3 case, a CP violating term arises from part of the Chern-Simons action
\bg
\int d^4 x d^2 y C_2 \wedge F \wedge F
\nd

Here the axion is given by 

\bg
a=\int d^2 y C_2
\nd

Again, we recognize the worldvolume instantons as being dissolved D1-brane instantons. The action of these instantons is proportional to  $e^{-\phi} \int B_2$ over the 2-cycle. This is again a modulus and the instanton contribution results in a coupling to this modulus of the form \eqref{ep} with $e^{-\phi}$ replaced by $\int B_2$ over the wrapped cycle. 

Similarly, we can consider D7 branes wrapping a 4-cycle. The axion will then be given by an integral of $C_4$ over that 4-cycle and the gauge instantons are dissolved Euclidean D3-instantons. The Kahler modulus that couples to the axion potential controls the size of this 4-cycle.

Finally we can also consider heterotic string theory. Here the coupling to the internal volume is manifest, as all the fields are spacetime fields and must be integrated over the full internal manifold when going to the 4D description.

Of course all of these scenarios can be related to each other by dualities, so it's not surprising that they lead us to the same conclusion. The common point illustrated in these examples is that upon coupling the field theory to gravity by embedding it in string theory, the gauge coupling becomes a dynamical variable itself. The instanton contributions don't just provide a potential for the axion, but couple the axion to a volume modulus and/or dilaton and the would-be local maximum in the axion potential still has a non-zero gradient along this direction satisfying the bound. 

The examples considered above are all supersymmetric and the Kahler and axion moduli are part of the same chiral multiplet in the effective 4D theory. The whole multiplet is stabilized and so the masses of the Kahler modulus and the axion are roughly the same. Indeed this is a problem one has to overcome when trying to obtain a QCD axion from string theory, as pointed out in \cite{conlon}. In a realistic model we want to eventually break supersymmetry anyway, so one may hope to use this breaking to avoid this problem. When supersymmetry is broken, it's possible to give each of these fields different masses and one might worry that this also breaks the above argument. Specifically one can imagine stabilizing the Kahler modulus at much higher energies than the axion and claim that the gauge coupling is effectively constant for all values of the axion, if we study the theory at low enough energies so that the Kahler modulus fluctuations are integrated out, and the effective theory only consists of the axion with a cosine potential. However, such reasoning is only valid in the absence of cross-terms between the heavy and light fields, and \eqref{ep} is precisely such a cross-term.

The idea of bypassing the gradient criterion by making the gauge coupling dynamical was already considered by the authors of \cite{murayama}. One of their proposals was to couple the quintessence field to the gluons, thus effectively making the gauge coupling depend on the quintessence field, and was rejected as a viable option for the QCD axion, since this would result in a coupling between the quintessence field and the nucleons, in tension with fifth-force constraints between nucleons. 

The situation we describe is similar in form, but differs in that the gluons don't couple to a separate light quintessence field. Instead the gluons are coupled to a much heavier field (the modulus that governs the coupling) that is not the same as the quintessence field considered in \cite{murayama}. Upon moving to the axion hilltop this modulus develops a runaway that respects the gradient criterion, but remains heavy, which suppresses the fifth-force interactions between nucleons.

One may worry that we are ignoring other contributions to the potential which in fact stabilize the Kahler moduli so that the total potential still violates the gradient criterion. However the instanton contribution considered above is precisely such a contribution, and flipping its sign will always convert a minimum to a runaway. The only way for this sign change to not affect the stability of the Kahler modulus is if it's stabilized by other more dominant effects to begin with.

\section{Kahler quintessence at axion hilltop?} \label{ht}

In the previous section we found that deviating far from the minimum of the axion potential leads to a runaway potential for the coupling modulus. Such potentials can be the candidates for a quintessence model that satisfies the swampland criteria. The Kahler modulus corresponding to the coupling constant which is part of the same multiplet as the QCD axion is not a good candidate for a quintessence field. On the other hand, we can consider other axions that don't couple strongly to the standard model sector, and their corresponding Kahler moduli can potentially serve as quintessence fields. In this section we will attempt to harness these dynamics to construct such a toy model of quintessence. We will find that the hilltop of a (non-QCD) axion potential generically contains runaway trajectories, although the slow-roll is bounded below. Such trajectories will generally be at high values of the potential and unstable in the axionic directions. To make the Kahler modulus a viable quintessence candidate, the value of the potential would have to be made small by extreme suppression of the one-loop determinant of the non-perturbative contribution to the superpotential. We will discuss the feasibility of this in string theory. We will also describe an attempt to stabilize this trajectory by introducing additional non-perturbative effects to create a dip in the axionic potential. This will naturally fix the slow-roll problem and produce a thawing hilltop quintessence model similar to \cite{susha}.

\subsection{Surveying the hilltop}

First we review the KKLT moduli stabilization procedure \cite{kklt}. Kahler moduli stabilization in KKLT-like scenarios is achieved by considering the non-perturbative brane-instanton and gaugino condensation effects. The resulting potential contains competing exponential terms with different exponents that balance each other out at some intermediate values of the volume. We take the effective superpotential and Kahler potential for a single Kahler modulus to have the following general form:

\bg \label{superpot}
W&=&W_0+ A \exp (-a T)  \nonumber \\
K &=& -3 \log (T + \bar T)
\nd

$W_0$ is independent of the Kahler modulus $T= X + i Y$, where $X$ could be the volume of some 4-cycle in the internal geometry, $Y$ is the axion corresponding to the 4-form threading that cycle, $a$ is of order $1/N$, where $N$ is the number of D7 branes on which the gaugino condensation takes place and generally depends on the other moduli as does $A$, which is a one-loop determinant that incorporates an integral over any instanton moduli other than spacetime translation modes.

In order to be able to ignore the higher order $\alpha^\prime$ corrections $X$ needs to be sufficiently large and we must check that this is the case at the end of our calculations.

To obtain the potential we need to compute
\bg
D_T W &=&-\frac{3}{2X}W_0 - (a+\frac{3}{2X}) A \exp(-a T), \nonumber \\
K^{T \bar{T}} &=& \frac{1}{3}X^2
\nd

\begin{figure}
\includegraphics[width=\columnwidth]{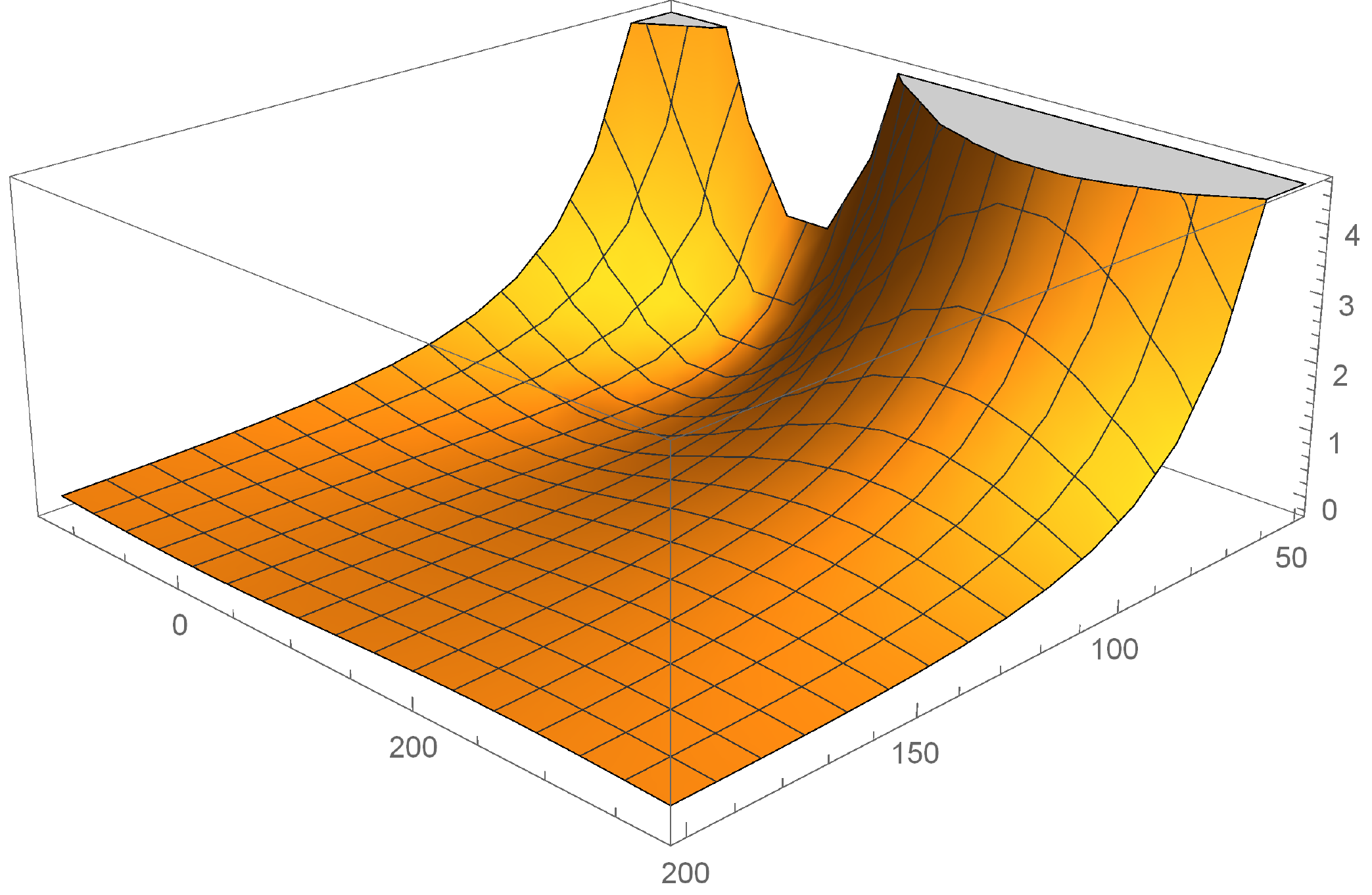} 
\caption{  \label{1np3d} The full scalar potential. The valleys contain the SUSY minimum, while the ridge is unstable. The vertical axis is scaled up to be of $\mathcal{O}(1)$. The scale is determined by the value of $A$.}
\end{figure}

This results in the following potential (fig. \ref{1np3d}).

\bg \label{potential}
V(X, Y) = \frac{ a e^{-2 a X} }{6X^2}\left(A^2(3 + aX) +3~ e^{a X} W_0 A~ \cos( a Y)\right) \nonumber \\
\nd

For a sypersymmetric minimum we need

\bg
DW = 0 \implies W_0 = - (\frac{2}{3}a X  +1) A  e^{-a (X+i Y)}, 
\nd
which will stabilize the axion at zero and the volume at some finite value. We will need this value to be large enough, so that the expansions in $X^{-1}$ make sense, but not so large that the KK modes become lighter than the Kahler modulus. At this minimum of the potential we have

\bg
V_c &=&- \frac{a^2 A^2 }{6 X_c} \exp(- 2 a X_c) 
\nd

The energy at the minimum is negative. At this point KKLT-type constructions invoke an uplift mechanism to obtain a meta-stable de Sitter minimum. We will take a different path and seek positive energy trajectories with sufficiently slow rolling to allow for at least one hubble time of expansion.

We consider setting the axion to $ a Y = \pi$. This will reverse the sign of the cosine term in \eqref{potential}. We have

\bg  \label{hilltop}
V(X, \pi/a)=\frac{a e^{-2 a X}}{6 X^2} \left(A^2(3 + a X) - 3 W_0 A e^{aX} \right)
\nd

Where we note that $W_0 A$ is negative, so the minus sign in front of the last term is misleading. The value of the potential is in fact positive definite and pushes the system to larger cycle volumes. This unstable configuration should be viewed as a non-equilibrium state in the same effective theory as the SUSY minimum. 

At this point we stop to comment on the constraints that the swampland conditions impose on our model. The canonically normalized fields are related to $X$ and $Y$ via
\bg
dx &=& \sqrt{3} \frac{dX}{X} \nonumber \\ 
dy &=& \sqrt{3} \frac{dY}{X} \\  
\nd

The distance criterion requires that our axion hilltop is within $\mathcal{O}(1)$ Planck units of the ground state as measured using the canonical field, which means we need

\bg \label{dist}
a X \gtrsim \mathcal{O}(1)
\nd
along our trajectory. This is also the condition for the 1-instanton contribution to the superpotential to be reliable.

We can check whether the potential \eqref{hilltop} satisfies the swampland criterion within this region. We find that near $X_c$

\bg \label{grad}
\frac{\left| \partial_x V \right|}{V} = \frac{2}{\sqrt{3}}\left(1+\frac{2}{3}a X_c \right) + \mathcal{O}(a (X-X_c)),
\nd
so the gradient criterion is satisfied for all values of $a X_c$.

If we start near $T = X_c + i \pi / a$, we expect the dynamics to initially roll toward larger $X$. However, the axion will tend to deviate from its hilltop value and roll down to zero. At this point the potential for $X$ will instead cause it to decrease down to the SUSY minimum. The rolling down of the axion will be accompanied by a transition from a positive energy density to a negative energy density. The duration of this process mainly depends on how quickly the axion leaves the hilltop and is therefore highly sensitive to initial conditions. The phenomenological viability of this model involves several considerations.

The first set of considerations, which is a general concern for all stringy quintessence models, comes from the requirement that the quintessence field is sufficiently ``dark'' to avoid tension with fifth-force constraints. Ensuring that our Kahler modulus doesn't couple to the Stanard Model fields is subtle \cite{dSvsQE}. The usual assumption is that we can insert the Standard Model into our compactification in such a way that the cycle controlled by our Kahler modulus doesn't couple to it. However, there is generically kinetic mixing between the various fields so this decoupling must be checked within specific models on a case-by-case basis.

The second set of phenomenological considerations concerns the tuning of initial conditions. Starting too close to the hilltop can lead to formation of domain walls, as different regions of space can roll off in different directions. The presence of such domain walls is problematic because they could be unstable, decompose and give rise to other axions. As pointed out in \cite{obied}, we must start sufficiently far from the hilltop so that quantum fluctuations don't lead to this domain wall formation, but sufficiently close so that we don't roll off along the axion directions too quickly. In \cite{obied} it was shown that this lower bound on the deviation from the hilltop is easy to obey and still leaves a range of initial conditions that remains near the hilltop for a Hubble time.

While suitable initial conditions exist, as with all ``hilltop'' models, there is a problem in justifying these initial conditions as the dynamics of the system naturally lead it away from them. One option would be to tunnel to them from a local minimum of similar energy, however demanding that the model respects the swampland criteria excludes this option. Another option is to modify the model so that this runaway trajectory was stabilized along the axionic direction. This will be the subject of the next subsection.

A third set of considerations concerns the value of the potential along our trajectory. The cosmological constant is famously exponentially small. The value of the potential at the axion hilltop is around 

\bg
V(X_c, \pi/a) = \left(\frac{8 a}{X_c^2} + \frac{4 a^2}{X_c}\right) A^2 e^{-2 a X_c} \sim \frac{A^2}{X_c^3}  
\nd

Note that since $a X$ is of order one, we can't exponentially suppress the value of the potential through the exponential terms. The inverse power law is too weak to achieve the exponentially small value required for today's vacuum energy, since taking $X_c \sim 10^{40}$ would make KK modes non-negligible. The only way this model could describe quintessence is if $A$ is exponentially small.

This requires some additional engineering, since $A$ arises as a one-loop determinant and is generally expected to be of order one. However, interactions with mobile D3 branes can alter this conclusion \cite{D3pot}. Indeed, in the presence of a mobile D3 brane, this one-loop determinant depends on the separation between the D3 and the stack of D7-branes wrapped on the supersymmetric cycle, on which the non-perturbative contribution arises. The result computed in \cite{D3pot} is

\bg \label{d3potential}
A = A_0 \left( f(w_i) \right)^{1/n},
\nd

Where $A_0$ is of order one, $w_i$ are the coordinates of the D3 brane and $f(y) = 0$ is the embedding equation for the supersymmetric cycle wrapped by $n$ D7 branes.
 
The dependence of $A$ on the position of the D3 also acts as a potential for the D3. This potential attracts the D3 toward the D7's. Unfortunately, this analysis also seems to suggest that there's no way of stabilizing the D3 at the required exponentially small distance without it being attracted to them and dissolving as worldvolume flux. However, this potential would also likely receive corrections at such small distances from the D7's. It is unclear what effect this has on the coefficient of the non-perturbative superpotential contributions, but note that if we are to believe eq \eqref{d3potential} all the way to zero separation, it would seem to imply that Kahler moduli stabilization is completely impossible in the presence of mobile D3 branes. Assuming this is not the case, this leaves one with some hope that an exponentially small one-loop determinant is achievable, but a detailed analysis of the relevant worldvolume dynamics and their backreaction would be required. For the remainder of this paper we will simply assume that such exponential suppression is possible.

Finally we must also make sure that the rolling along our trajectory is sufficiently slow. While not as constrained as inflation models, which requires many $e$-folds of slow-rolling expansion, a viable quintessence model still requires a relatively small slow-roll parameter, which means the gradient criterion needs to be nearly saturated. Indeed, if the field excursions are limited to $\mathcal{O}(1)$ Planck units, then to obtain at least one Hubble time of expansion, the gradient slow-roll parameter must be around $\mathcal{O}(1)$ as well, since the number of $e$-folds of expansion is proportional to 
\bg
N\propto\int \frac{V}{|\nabla V|} dx,
\nd
where $x$ is the canonically normalized field. We also saw that as \eqref{grad} approaches saturation of the gradient criterion we also start to approach the limits allowed by the distance criterion \eqref{dist} along the axionic direction. This non-saturation of the gradient criterion is not hierarchically large, but in the light of recent work assessing the compatibility of the gradient criterion with present observational constraints\cite{phenomquint}, this model's potential is too steep for even the most optimistic estimates.

To recapitulate, we have studied the KKLT potential at an axionic maximum as a potential candidate quintessence model. We discussed potential ways to address the requirements that our quintessence field is sufficiently ``dark'' and that the potential is exponentially small. These issues must be addressed at the level of specific stringy constructions and we will henceforth assume that suitable compactifications and brane configurations exist. We also found constraints imposed on the parameters of the model by the swampland criteria. Satisfying the distance criterion along the axionic direction guarantees that we satisfy the gradient criterion along the Kahler modulus direction. In fact the potential along the Kahler direction is too steep for phenomenological viability, but not hierarchically so. The model also suffers from an initial conditions fine-tuning problem, since the desired trajectory lies at the axionic hilltop.

In the next subsection we will modify this model by adding a second non-perturbative contribution, which will simultaneously solve the slow-roll problem and improve the fine-tuning problem.

\subsection{Carving out a valley}

To bypass the initial conditions problem, we can attempt to modify the model to create a local minimum along the axionic direction at positive energy. In general, the number of non-perturbative effects in a given compactification will exceed the number of axion fields, and so the potential will generally admit local minima along the axionic directions. The authors of \cite{aoe} have devised a powerful framework for systematically analyzing the potential for a large numbers of axions with non-perturbative cosine potentials and local meta-stable minima are a generic feature of such axionic potentials. In these works they made no attempt to embed the axions in string theory so they had no reason to consider a dynamical coupling. As such, these potentials would seem to violate the gradient criterion in the same way that the Peccei-Quinn mechanism did, but are consistent with the refined version. Moreover, as we argued in section \ref{pq}, it's impossible to ignore the modulus that governs the coupling far from the global minimum of the axion potential and so we generically expect to obtain arunaway potential for the coupling, and the swampland criteria have to be considered in the full field space.

If an axionic local minimum happens at positive energy, it could create a valley with runaway behavior along the $X$ direction, providing a stable trajectory for a quintessence model.\footnote{\cite{dSvsQE} considers models with similar runaway ``valleys'' and point out problems related to strong quantum corrections to the potential after supersymmetry breaking. Since our EFT is ultimately supersymmetric, we can expect to have control over the quantum corrections to the scalar potential, avoiding these problems.} The walls of this valley could either then shallow out at sufficiently large $X$ allowing the axion $Y$ to roll back down to its global minimum, or the system could tunnel out of the valley and roll back to the SUSY minimum. This way the dynamics would naturally justify initial conditions away from the global minimum. We must then look for values of the parameters that allow for a sufficiently long lasting slow-roll period at exponentially suppressed energy\footnote{As in the previous subsection, the exponentially low energy would have to be achieved by an exponential suppression of the appropriate one-loop determinants.} and end in an eventual decay into the SUSY AdS minimum. The dynamics in such a model would never need to leave the regime of validity of the EFT describing the SUSY minimum.

To obtain local minima in the axion potential we need more non-perturbative contributions than axions. The simplest realization of this is the well-studied ``racetrack'' scenario \cite{racetrack}, where the superpotential is now given by

\bg \label{rt}
W&=&W_0+ A \exp (-a T) + B \exp(-b T) \nonumber \\
\nd

The resulting scalar potential is

\bg
V(&X&, Y)=\frac{e^{-(a+b)X}}{6 X^2} \left( a A^2 (aX+3) e^{(b-a)X} \right. \\
&+& b B^2 (bX+3) e^{(a-b)X}  \nonumber \\
&+&AB(2abX+3a+3b) \cos[(a-b) Y] \nonumber \\
 &+& \left. 3 W_0 ( a A e^{bX}\cos[aY] + bB e^{aX} \cos[bY]) \right) \nonumber
\nd

Similar to the previous section, the distance criterion requires at least one of the following:
\bg
a X \gtrsim 1 \qquad b X \gtrsim 1 \qquad (a-b) X \gtrsim 1,
\nd
to hold, depending on which cosine term is dominant, otherwise neighboring local axionic minima will be more than at a superplanckian distance away from the absolute minimum and we can no longer trust the effective field theory at one minimum to describe the other. This would also constitute a violation of the weak gravity conjecture as described in \cite{moritz}.

The absolute minimum of the potential $T = X_c + 0 i$ can once again be found by demanding $DW=0$, which leads to

\bg \label{critW}
W_0 = -(1+\frac{2}{3}a X_c) e^{-a X_c} - (1+\frac{2}{3} b X_c) e^{-b X_c} 
\nd

To stay within the regime of validity of the same EFT that describes this minimum, $X$ needs to remain within a factor of $e^{1/\sqrt{3}}$ of $X_c$. If we fix $X = X_c$ and study the potential along the axionic directions we obtain

\bg \label{ypotential}
V(&X&_c, Y)=\frac{e^{-(a+b)X_c}}{6X_c^2} \left( a A^2 (aX_c+3) e^{(b-a)X_c} \right.  \\
&+& b B^2 (bX_c+3) e^{(a-b)X_c} \nonumber \\
&+&AB(2abX_c+3a+3b) \cos[(a-b) Y] \nonumber \\
&-&( a A^2 (2a X_c+3) e^{(b-a)X_c}+a A B (2bX_c+3))  \cos [a Y] \nonumber \\
&-&( b B^2 (2 b X_c +3) e^{(a-b) X_c} + b A B(2 a X_c + 3) ) \cos[ b Y] \left. \right) \nonumber
\nd

In principle we have enough freedom in the parameters of the above potential to set the coefficients of any of the cosines as well as the $Y$-independent term to anything we want. It is therefore not too difficult to construct a potential that would have a local minimum along the $Y$-direction near $X=X_c$. 

A simple and perhaps most obvious choice of parameters that yields a local axionic minimum is 

\bg \label{attempt1}
b=2a \qquad A=B
\nd

\begin{figure}

\includegraphics[width=\columnwidth]{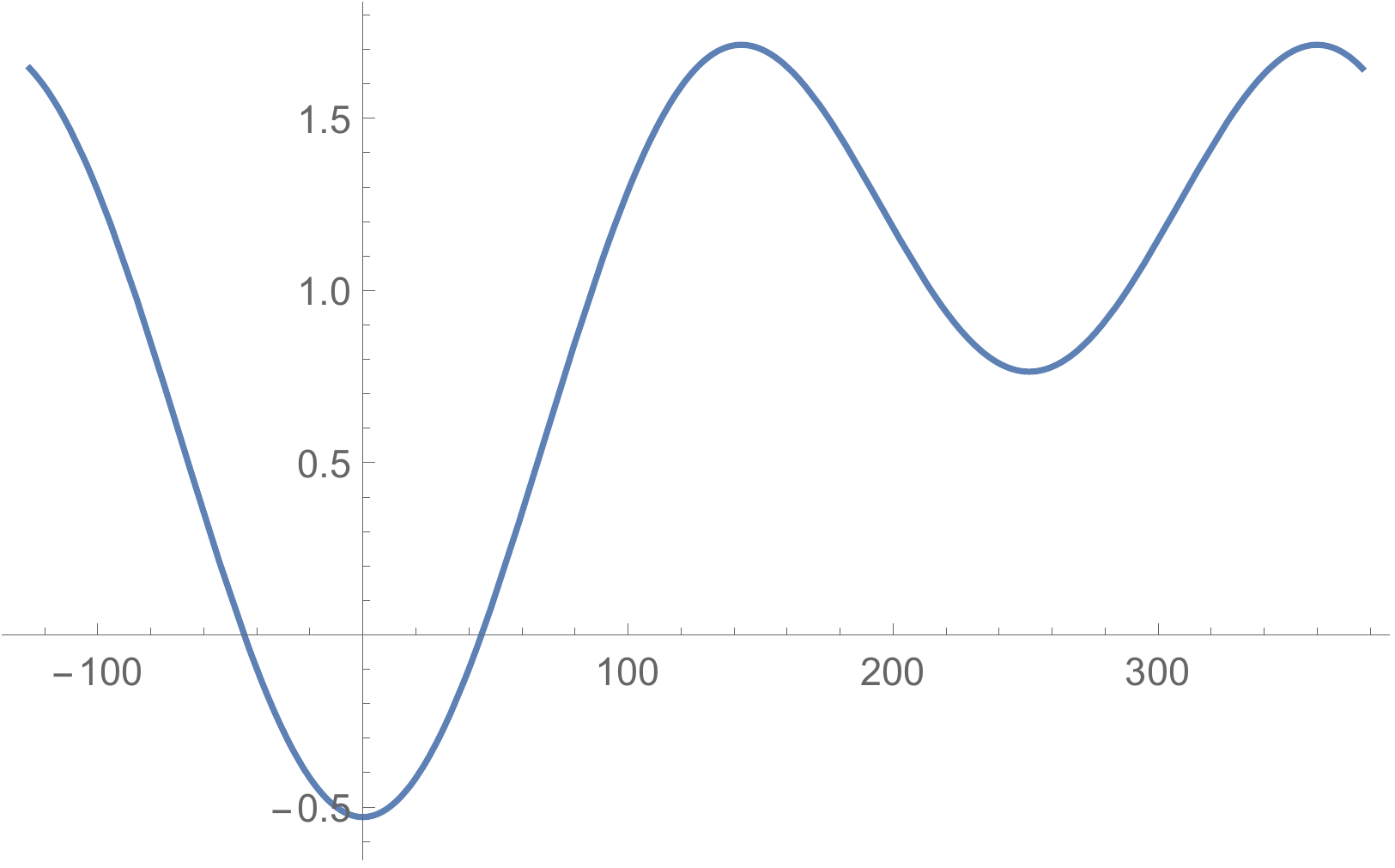} 
\caption{  \label{localmin} The axion potential at $X=X_c=100$ with parameters $a=1.25, b=2.5, A=B$ chosen to give a local minimum. The vertical axis is scaled up to be of $\mathcal{O}(1)$. The real scale is determined by the values of $A, B$}
\end{figure}

\begin{figure}

\includegraphics[width=\columnwidth]{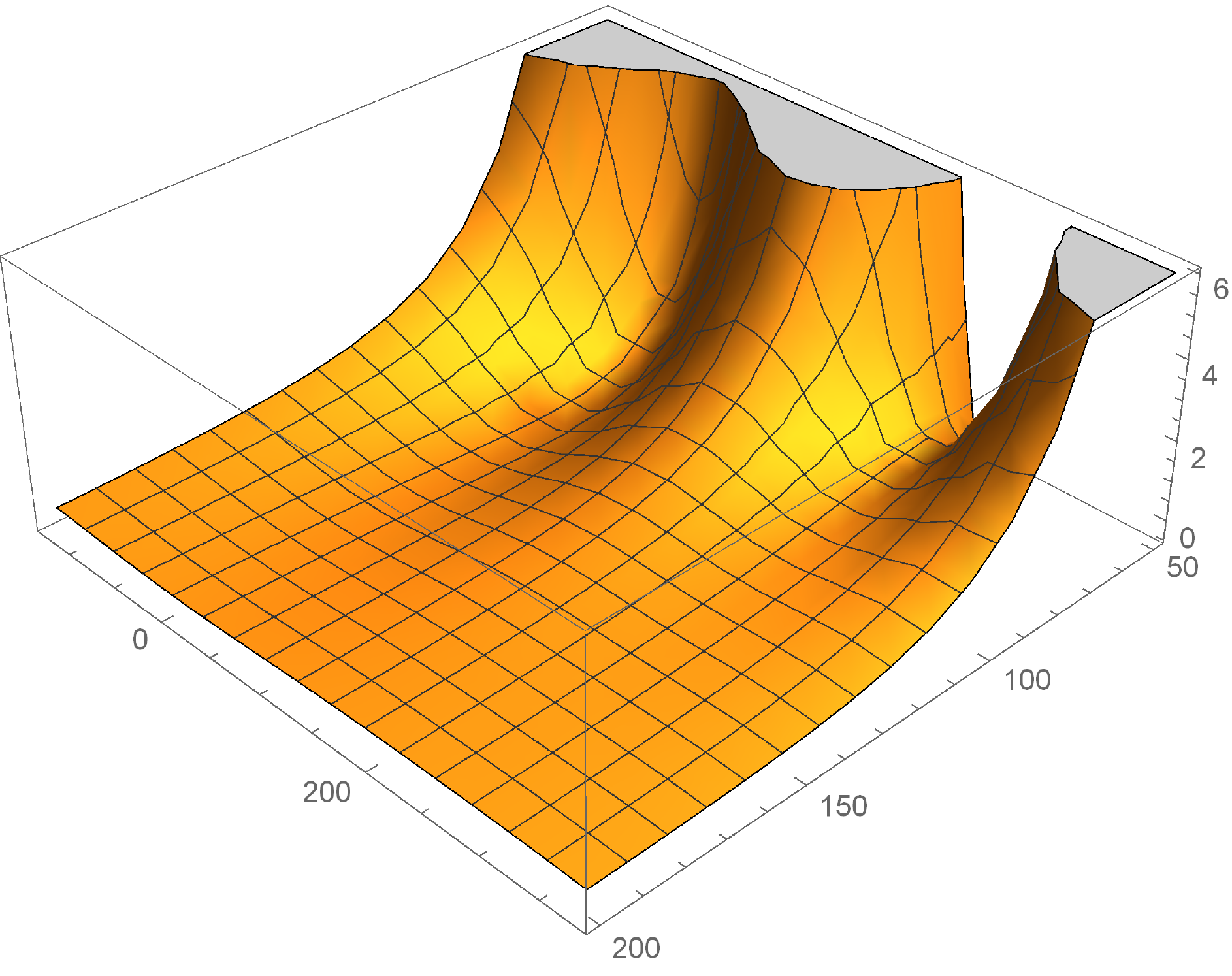} 
\caption{  \label{2np3d} The full scalar potential with two non-perturbative effects with $a=1.25, b=2.5, A=B$. The deep valley contains the SUSY minimum, and the shallower valley contains the saddle point. Starting close enough to the saddle allows for several e-folds of expansion, while the tail of the runaway remains too steep. The vertical axis is scaled up to be of $\mathcal{O}(1)$.}
\end{figure}

Plotting the scalar potential with such a choice of parameters (fig. \ref{localmin} and \ref{2np3d}) we clearly get the valley that we seek.

\begin{figure}
\includegraphics[width=\columnwidth]{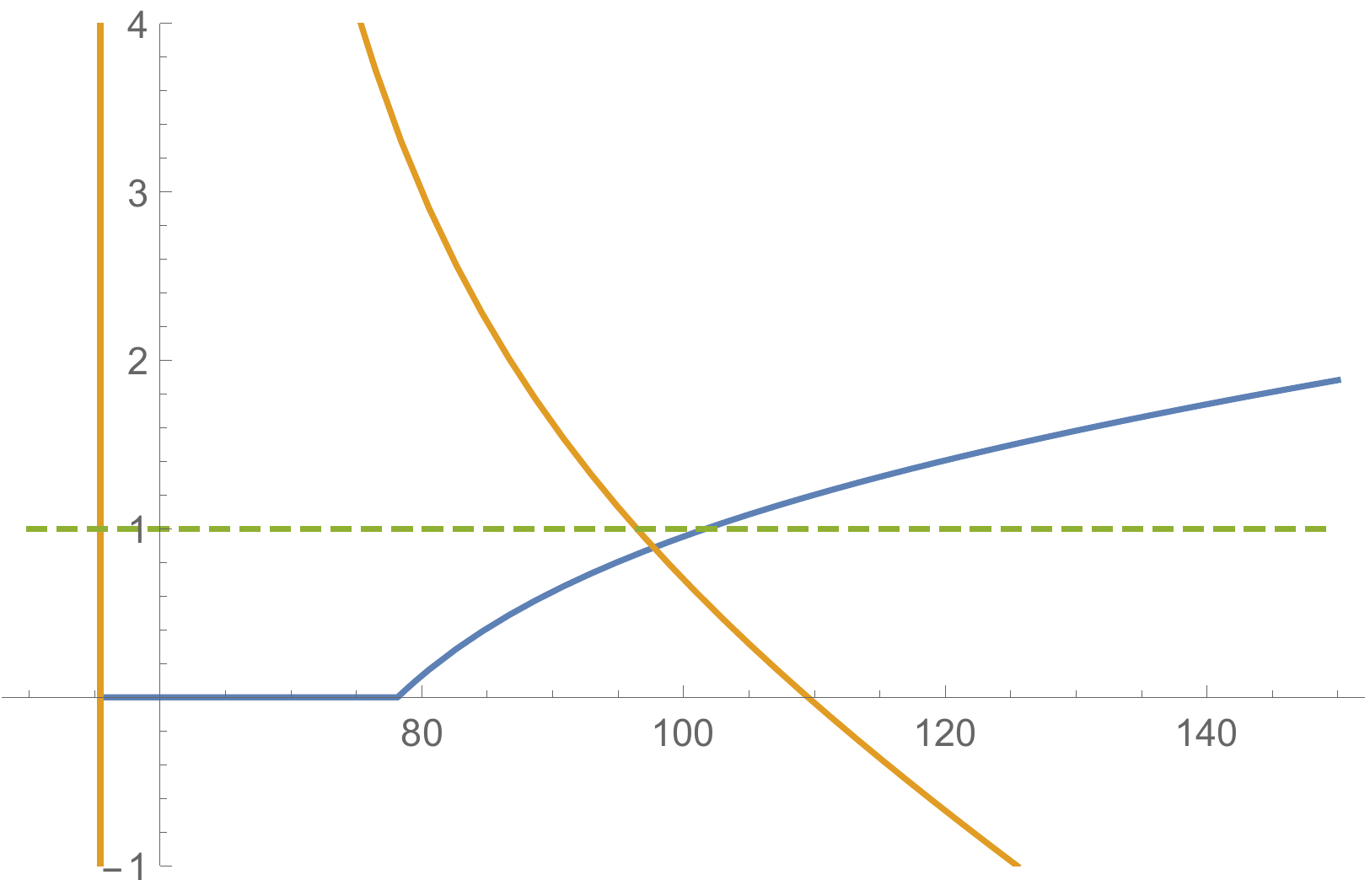} 
\caption{  \label{criteriaplot} The quantities $\tilde{c}$ and $\tilde{a}$ appearing in the gradient and hessian swampland criteria. At any value of $X$ at least one of the criteria is satisfied. However sufficient slow-roll for an Hubble time of expansion only occurs in the region satisfied by the hessian criterion.}
\end{figure}

\pagebreak
The potential, however, also develops a new feature in the form of a local maximum along the $X$ direction. This maximum obviously violates the gradient criterion and forces us to consider the hessian criterion, which we originally intended to avoid making use of. Figure \ref{criteriaplot} plots the two quantities appearing in the gradient and hessian criteria respectively.

\bg
\tilde{c} = \frac{\partial_x V}{V}; \qquad
\tilde{a} = -\frac{\partial_x^2 V}{V}
\nd

\begin{figure}
\includegraphics[width=\columnwidth]{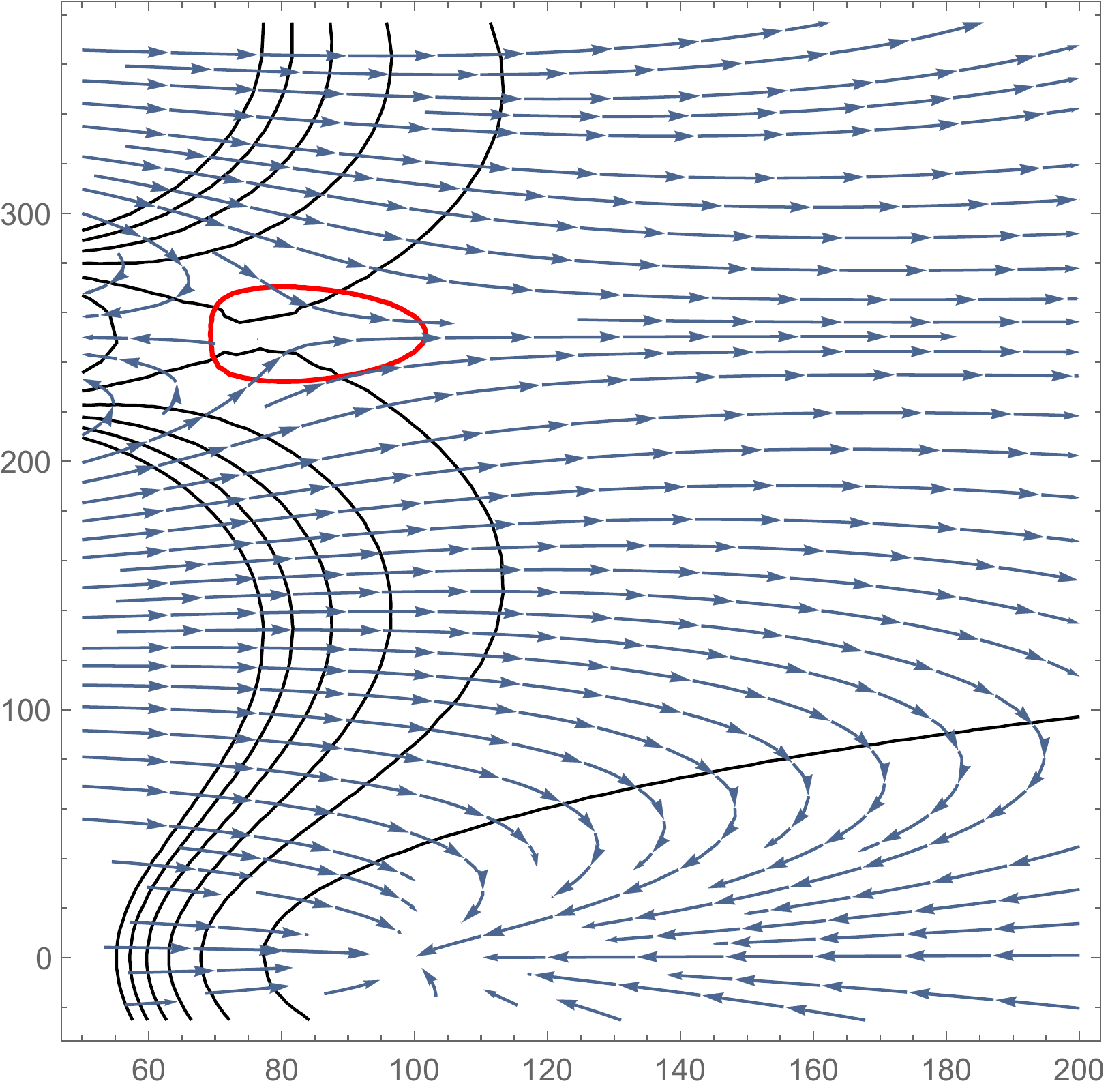} 
\caption{  \label{2npstream} Contour plot of the potential with $A=B, a=1.25, b=2.5, X_c= 100$ with gradient streamlines. The red contour denotes the region where the gradient criterion is saturated. Near the saddle, the slow-roll parameter is small enough to generate several Hubble times of expansion.}
\end{figure}

We see a crossover between the two quantities, such that as one criterion starts to be violated the other becomes satisfied. We note however that in the region where the gradient criterion is satisfied, $\tilde{c}$ rather quickly goes up above 1. This means that the tail of the potential is unsuitable for a slow-rolling quintessence model just as we had in the one non-perturbative effect case. Indeed one can check for generic values of the parameters that if we wish to have even one $e$-fold of expansion by the time the field reaches the criterion crossover point the starting point must be fairly close to the hilltop, where most of the expansion takes place. This means that this potential only acts as a quintessence model in the region roughly between the saddle point and the crossover, transforming our model into a thawing quintessence model.\footnote{In this way this model is similar to that in \cite{susha}, except our model also has a non-trivial axion potential.} Thus it seems we have traded one initial conditions problem for another. However note that the space of suitable initial conditions is larger than just the neighborhood of the saddle, as there are clearly some trajectories that roll \emph{into} the slow-roll region (fig. \ref{2npstream}).

One may consider expanding the model by including additional axions with more non-perturbative effects along the lines of \cite{hristov}. Interestingly, it seems this only makes it more difficult to achieve positive energy local minima in the axion directions. Indeed, in the second paper of \cite{aoe} the authors determine the maximum energy for local minima in generic multi-axion landscapes and find that it goes down with the number of axions. This means that obtaining a positive energy valley for the Kahler modulus to roll down becomes increasingly difficult as the number of axions increases.

\section{Conclusion}

The swampland criteria aim to articulate restrictions on effective field theories that can allow one to determine from a dimensionally reduced perspective whether a given model can arise from a string theory compactification. In this work we have examined the interplay between the gradient, distance and hessian swampland criteria and the non-perturbative axion potentials that generically arise in string compactifications. First we pointed out that the tensions between local maxima of axion potentials and the gradient criterion discussed in \cite{murayama} are in fact not there if one properly considers the coupling parameter as a dynamical variable. Moving to hilltops in the axion potential generally destabilizes the real parts of the Kahler moduli and the potential has a runaway behavior in that direction. We note that if moving to the hilltop doesn't violate the distance criterion, this is generally accompanied by the runaway behavior also satisfying the gradient criterion.

Runaway potentials may be viable ingredients for quintessence models, and in the rest of the paper we explore the possibility of constructing such a model within the KKLT moduli stabilization framework. Rather than using a supersymmetry-breaking uplifting ingredient, such as anti-branes, we proposed to view the current state of the universe as a non-equilibrium, non-supersymmetric positive energy state in the same supersymmetric effective field theory as the KKLT AdS minimum. In this non-equilibrium state the axion is at its hilltop far from the global minimum and the destabilized Kahler modulus acts as a quintessence field. 

We discussed the restrictions that the swampland criterial place on the space of parameters and searched for initial conditions that could ensure a suitably long period of exponentially small, nearly constant positive energy density, without violating the distance criterion. One of the parameters that required extreme fine-tuning is the coefficient of the non-perturbative contribution to the superpotential, which is given by a one-loop determinant. This coefficient effectively controls the height of the potential, and therefore needs to be exponentially small. The only way we are aware of tuning this one-loop determinant is via backreaction of mobile D3 branes \cite{D3pot}, however this results in an attractive potential between the D3's to the D7's and a more detailed analysis of what happens at small separation is required. 

In addition to tuning the height of the potential, this model is unstable in the axionic direction requiring fine-tuning of initial conditions. A further problem was the fact that the hilltop trajectory is too steep, although not hierarchically so, to allow one Hubble time of expansion without violating the distance criterion.

We attempted to remedy these problems by turning the axionic hilltop into a local valley, stabilizing the desired trajectory, by considering a ``racetrack'' scenario with two non-perturbative contributions to the superpotential. This allowed us to generate a local minimum along the axionic directions, creating a runaway ``valley''. This also generated a local maximum along the Kahler modulus direction resulting in a saddle point of the full potential. As we move away from the saddle in the positive Kahler modulus direction, the hessian criterion gives way to the gradient criterion, with at least one being satisfied along the full valley. Achieving one full Hubble time of expansion still requires spending some time near the saddle point,transforming the model into a thawing quintessence model. However, this does not mean one needs to arbitrarily start at the saddle point, as there are trajectories which roll down towards it.

\section*{Acknowledgements}

We would like to thank Keshav Dasgupta, Robert Brandengerger, Susha Parameswaran, Ed Hardy and Jim Cline for helpful discussions.
The work of ME is supported in part by the Natural Sciences and Engineering Research Council of Canada. 

{}
 \end{document}